\documentstyle[12pt]{article}
\begin{document}
\begin{titlepage}
\null

\begin{flushright}
NBI-HE-93-49\\
KOBE-TH-93-06\\
August 1993
\end{flushright}

\vspace{7mm}
\begin{center}
  {\Large\bf Toroidal Orbifold Models \par}
  {\Large\bf with a Wess-Zumino Term\par}
  \vspace{1.5cm}
  \baselineskip=7mm

  {\large Jens Ole Madsen\footnote{Address until Feb. 28, 1994 :
LAPP, B.P. 110, Chemin de Bellevue, F-74941 Annecy-le-vieux, France.} \par}
\vspace{5mm}
  {\sl The Niels Bohr Institute, University of Copenhagen\\
     Blegdamsvej 17, DK-2100 Copenhagen \O, Denmark\par}

\vspace{7mm}
  {\large and \par}
\vspace{7mm}

   {\large Makoto Sakamoto \par}
\vspace{5mm}
     {\sl Department of Physics, Kobe University\\
      Rokkodai, Nada, Kobe 657, Japan \par}

\vspace{3cm}

{\large\bf Abstract}
\end{center}
\par

Closed bosonic string theory on toroidal orbifolds is studied
in a Lagrangian path integral formulation.
It is shown that a level one twisted WZW action whose field
value is restricted to Cartan subgroups of simply-laced Lie
groups on a Riemann surface is a natural and nontrivial
extension of a first quantized action of string theory on
orbifolds with an antisymmetric background field.

\end{titlepage}
\setcounter{footnote}{0}
\baselineskip=7mm


String theory on toroidal orbifolds \cite{orb} has been
studied from both operator formalism and path  integral
formalism points of view.
Some of the advantages of the operator formalism are that the
algebraic structure is clear and that it is  possible to
formulate the theory without Lagrangians or actions.
On the other hand, in the Lagrangian path integral formalism
the geometrical or topological structure is transparent and the
generalization to higher genus Riemann surfaces is obvious.
The interrelation between the two formalisms is not, however, trivial.

The purpose of this paper is to study toroidal
orbifold models with
nontrivial twists in the Lagrangian path integral formalism and
to clarify the topological structure of the orbifold models.
In the operator formalism of closed bosonic string theory,
we can introduce a left- and right-moving coordinate
$(X^I_L,X^I_R)$.
An orbifold is obtained by dividing a torus by the action of a
discrete symmetry group $P$ of the torus.
Any element of $P$ can in general be represented by a rotation $U$
and a shift $v$ (for symmetric orbifolds) \cite{orb}.
On the orbifold a point $(X^I_L,X^I_R)$ is identified with
$(U^{IJ}X^J_L + 2\pi v^I, U^{IJ}X^J_R - 2\pi v^I)$ for all
$(U,v) \in P.$
If we wish to formulate the orbifold model in the Lagrangian
path integral formalism, the following two problems arise:
In the path integral formalism, a one-loop vacuum amplitude
is given by the
functional integral \cite{Polyakov},
\begin{equation}
\int_\Sigma \frac{[dg_{\alpha \beta}] [dX^I]}{V}
\ \ \exp \{ -S[X,g] \}\ ,
\label{vac ampli}
\end{equation}
where $g_{\alpha \beta}$ is a metric of a Riemann surface $\Sigma$
of genus one and $X^I$ is a string coordinate, which maps $\Sigma$
into a target space.
The $V$ is a volume of local symmetry groups.
The action $S[X,g]$ would be of the from,
\begin{equation}
S[X,g]\ =\ \int^1_0 d^2\sigma \frac{1}{2\pi} \left\{
         \sqrt{g} g^{\alpha \beta} \partial_\alpha X^I
         \partial_\beta X^I\
         -\ iB^{IJ}\varepsilon^{\alpha \beta}
         \partial_\alpha X^I \partial_\beta X^J\ \right\}\ ,
\label{action}
\end{equation}
where $B^{IJ}$ is an antisymmetric constant background field, which
has been introduced by Narain, Sarmadi and Witten \cite{N-S-W} to
explain Narain torus compactification \cite{Narain} in the
conventional approach.
The first problem is that the $B^{IJ}$-term in the action
(\ref{action}) becomes ill-defined for twisted strings associated
with twists $U^{IJ}$ which do not commute with $B^{IJ}$
\cite{I-N-T}.
In ref.\cite{top}, orbifold models with such twists have been
studied in the operator formalism in detail.
It has been shown that the orbifold models with nonvanishing
$[B,U]$ for some $U$ exhibits various anomalous behavior.
The analysis has suggested that those orbifold models are
topologically quite different from orbifold models with vanishing
$[B,U]$ for all $U.$
However, the topological structure has not clearly been understood.
The second problem is that a combination
$X^I=\frac{1}{2}(X^I_L+X^I_R)$ appears in the action (\ref{action})
but a combination $\frac{1}{2}(X^I_L-X^I_R)$ does not.
Hence, it seems that there is no way to impose the twisted
boundary condition corresponding to the identification
$(X^I_L,X^I_R) \sim (U^{IJ}X^J_L+2\pi v^I, U^{IJ}X^J_R-2\pi v^I)$
unless $v^I=0$ or unless we introduce new degrees of freedom
corresponding to $\frac{1}{2}(X^I_L-X^I_R)$ besides $X^I.$
In this paper we shall propose a solution to the two problems and show that
the action (\ref{action}) should be replaced by a WZW action for
orbifold models with nontrivial twists.

Let us discuss the first problem mentioned above.
To simplify our discussion, we will consider orbifold models
without shifts ($v=0).$
A $D$-dimensional torus $T^D$ is defined by identifying a point
$\{ X^I\}$ with $\{ X^I+\pi w^I\}$ for all $w^I \in \Lambda,$
where $\Lambda$ is a $D$-dimensional lattice.
An orbifold is obtained by dividing the torus by the action of
a discrete symmetry group $P$ of the torus, i.e., a point
$\{ X^I\}$ is identified with $\{U^{IJ}X^J+\pi w^I\}$
for all $U \in P$ and $w \in \Lambda$ on the orbifold.
We will here call a twist $U$ topologically trivial (nontrivial)
if $[B,U]=0 \ ([B,U]\not= 0).$

Throughout this paper, we will restrict our attention to a class
of the following orbifolds: The lattice $\Lambda$ is taken to be
a root lattice $\Lambda_R(\cal G)$ of a simply-laced Lie algebra
$\cal G$ with rank $D$ and the squared length of the root vectors
is normalized to two.
In this normalization the weight lattice $\Lambda_W(\cal G)$ is
just the dual lattice of $\Lambda_R(\cal G)$.
The antisymmetric background field $B^{IJ}$ is given through the
relation,
\begin{equation}
\alpha^I_i B^{IJ} \alpha^J_j\ =\ \alpha^I_i \alpha^I_j\qquad
  \mbox{mod\ 2}\ ,
\label{def B}
\end{equation}
where $\alpha_i$ is a simple root of $\cal G$
\footnote{The above choice of $\Lambda$ and $B^{IJ}$ leads to
the $(D+D)$-dimensional Lorentzian even self-dual lattices
introduced by Englert and Neveu \cite{E-N}.}.
The rotation matrices $U^{IJ}$ are chosen to be automorphisms
of $\Lambda_R(\cal G)$.
Then, we find that $U^{IJ}$ do not always commute with $B^{IJ}$
and that they satisfy $(B-U^TBU)^{IJ}w^J \in 2\Lambda_W(\cal G)$
for all $w^I \in \Lambda_R(\cal G)$, which is a necessary
condition to construct consistent orbifold models in the operator
formalism \cite{top}.
We note that (twisted) affine Ka\v c-Moody algebras are realized
in these orbifold models through the Frenkel-Ka\v c-Segal mechanism
\cite{F-K-S}.
Since strings propagate on the orbifolds, the string coordinate
in general satisfies the following twisted boundary condition:
\begin{eqnarray}
X^I(\sigma^1+1,\sigma^2) &=& U^{IJ} X^J(\sigma^1,\sigma^2) +
                             \pi w^I\ ,\nonumber\\
X^I(\sigma^1,\sigma^2+1) &=&
                \widetilde{U}^{IJ} X^J(\sigma^1,\sigma^2) +
                             \pi \widetilde{w}^I\ ,
\label{twist bc}
\end{eqnarray}
for some $U,\widetilde{U} \in P$ and $w,\widetilde{w} \in \Lambda$.
The consistency of the above boundary condition requires
\begin{eqnarray}
[U,\widetilde{U}] &=& 0\ ,\nonumber\\
(1-\widetilde{U})^{IJ}w^J &=& (1-U)^{IJ}\widetilde{w}^J\ .
\label{consist}
\end{eqnarray}
One might expect that the action for the twisted string satisfying
eq.(\ref{twist bc}) is given by eq.(\ref{action}).
{}For trivial twists $U$ and $\widetilde{U}$, i.e.,
$[B,U] = [B,\widetilde{U}]
= 0$, this is true.
{}For nontrivial twists, the noncommutativity of $B^{IJ}$ and $U^{IJ}$
(or $\widetilde{U}^{IJ})$, however, causes trouble because the
$B^{IJ}$-term in eq.(\ref{action}) is not well-defined.
We wish to find a generalization of the $B^{IJ}$-term, which must be
well-defined even for nontrivial twists.
Since the $B^{IJ}$-term is independent
of the metric  $g_{\alpha \beta}$, the generalization of the
$B^{IJ}$-term will also
be a topological term independent of $g_{\alpha \beta}$.
It seems that there is no such desired term in two dimensions.
A key observation to solve our problem is that the $B^{IJ}$-term
can be rewritten as a \lq\lq truncated" Wess-Zumino term modulo $2\pi i$
for strings on tori \cite{C-G}.
According to the prescription of ref.\cite{C-G}, let us introduce
a field $\phi(\sigma^1,\sigma^2)$ defined by
\begin{equation}
\phi(\sigma^1,\sigma^2)\ =\ \exp \left\{
         i2X^I(\sigma^1,\sigma^2) H^I \right\}\ ,
\label{def phi}
\end{equation}
where $H^I$ is a generator of the Cartan subalgebra of $\cal G$
and is normalized such that $Tr(H^IH^J)=\delta^{IJ}$.
We note that $\phi(\sigma^1,\sigma^2)$ is a mapping from $\Sigma$
into the Cartan subgroup of the group $G$, the algebra of which is $\cal G$.
It is easy to see that the first term in eq.(\ref{action})
can be rewritten as
\begin{equation}
-\frac{1}{8\pi} \int^1_0 d^2\sigma \sqrt{g} g^{\alpha \beta}
      \, Tr\left(\ \phi^{-1}\partial_\alpha \phi\, \phi^{-1}
      \partial_\beta \phi\ \right)\ .
\end{equation}
A Wess-Zumino term \cite{WZ}-\cite{Petersen} at level one is given by
\footnote{For some orbifold models, the Wess-Zumino term defined in
eq.(\ref{def WZW}) might be modified to make it well-defined
\cite{F-G-K}. We will not discuss this problem in this
paper.}
\begin{equation}
\Gamma_{WZ}( \widetilde{\phi})\ =\
    -\frac{i}{12\pi} \int_M Tr\left( \widetilde{\phi}^{-1}
      d\widetilde{\phi} \right)^3\ ,
\label{def WZW}
\end{equation}
where $M$ is a three dimensional manifold whose boundary is $\Sigma$
and $\phi$ is extended to a mapping $\widetilde{\phi}$ from
$M$ into $G$
with $\widetilde{\phi}\big{\vert}_\Sigma = \phi$.
It has been shown in ref.\cite{C-G} that for strings on tori the
Wess-Zumino term (\ref{def WZW}) is equivalent to the $B^{IJ}$-term in eq.
(\ref{action}) modulo $2\pi i$ through the relation (\ref{def phi}).
We shall show that $\Gamma_{WZ}$ is just the term we wish to find, i.e.,
$\Gamma_{WZ}$ is well-defined even for topologically nontrivial
twisted strings and is reduced to the $B^{IJ}$-term for
topologically trivial ones.

In order to determine what boundary condition we should impose
on the field $\widetilde{\phi}$, let us briefly review automorphisms
of Lie algebras.
Let $\cal G$ be a simply-laced Lie algebra.
We normalize the squared length of the root vectors to two.
In the Cartan-Weyl basis, the algebra $\cal G$ is given by
\begin{eqnarray}
& &[H^I,H^J]\ =\  0\ ,\nonumber\\
& &[H^I,E^\alpha ]\ =\ \alpha^I E^\alpha\ ,\nonumber\\
& &[E^\alpha,E^\beta ]\ =\ \cases{
     \alpha^I H^I\ ,&\quad\mbox{for $\alpha+\beta=0$}\ ,\cr
     \varepsilon(\alpha,\beta) E^{\alpha+\beta}\ ,&\quad
     \mbox{for $\alpha+\beta=$ a root vector}\ ,\cr
     0\ , &\quad\mbox{otherwise}\ ,\cr}
\label{algebra}
\end{eqnarray}
where $H^I$ is a generator of the Cartan subalgebra and $E^\alpha$
is a step operator associated with the root vector $\alpha$.
By suitably choosing phases of the step operators, the structure
constant $\varepsilon(\alpha,\beta)$ may be given by
\cite{IKKS}
\begin{equation}
\varepsilon(\alpha,\beta)\ =\ \exp \left\{-i\frac{\pi}{2}
     \alpha^IB^{IJ}\beta^J\ \right\}\ ,
\label{epsilon}
\end{equation}
where $B^{IJ}$ is defined in eq.(\ref{def B}).
We consider automorphisms of the algebra $\cal G$ given by
\begin{eqnarray}
\tau(H^I) &=& (U^T)^{IJ}H^J\ ,\nonumber\\
\tau(E^\alpha) &=& \eta(U;\alpha) E^{U\alpha}\ ,
\label{auto}
\end{eqnarray}
where
\begin{equation}
\eta(U;\alpha)\ =\ \exp \left\{ i\frac{\pi}{2} \alpha^I
      C_U^{IJ}\alpha^J \right\}\ .
\label{eta}
\end{equation}
The $C^{IJ}_U$ is a symmetric matrix defined through the relation,
\begin{equation}
\alpha^I_i C^{IJ}_U \alpha^J_j\ =\ \frac{1}{2} \alpha^I_i
     (B-U^TBU)^{IJ} \alpha^J_j\qquad \mbox{mod\ 2}\ .
\label{def C}
\end{equation}
It should be noted that for nontrivial twists $U$, setting
$\eta(U;\alpha)$ equal to one is incompatible with the invariance of the
structure constants under the automorphism $\tau$.

Let us now return to our problem.
We wish to find a boundary condition which assures the
single-valuedness of $Tr(\widetilde{\phi}^{-1}d\widetilde{\phi})^3$
on $M$.
Since $\widetilde{\phi}^{-1}d\widetilde{\phi}$ is a one form with values in
the Lie algebra $\cal G$ and is equal to $\phi^{-1}d\phi$ on
$\partial M=\Sigma$, $Tr(\widetilde{\phi}^{-1}d\widetilde{\phi})^3$
will be single-valued on $M$ if
$\widetilde{\phi}^{-1}d\widetilde{\phi}$
obeys the following boundary condition:
\begin{eqnarray}
\widetilde{\phi}^{-1}d\widetilde{\phi}(t,\sigma^1+1,\sigma^2) &=&
    \widetilde{\phi}^{-1}d\widetilde{\phi}(t,\sigma^1,\sigma^2)
    \Big\vert_{H^I\rightarrow (U^T)^{IJ}H^J
     \atop E^\alpha\rightarrow \eta(U;\alpha)E^{U\alpha}}\ ,
     \nonumber\\
\widetilde{\phi}^{-1}d\widetilde{\phi}(t,\sigma^1,\sigma^2+1) &=&
    \widetilde{\phi}^{-1}d\widetilde{\phi}(t,\sigma^1,\sigma^2)
    \Big\vert_{H^I\rightarrow (\widetilde{U}^T)^{IJ}H^J
     \atop E^\alpha\rightarrow \eta(\widetilde{U};\alpha)
     E^{\widetilde{U}\alpha}}\ ,
\label{bc tildephi}
\end{eqnarray}
where $t,\sigma^1$ and $\sigma^2$ are coordinates for $M.$

The Wess-Zumino term is independent of the metric $g_{\alpha \beta}$
and vanishes for any infinitesimal variation, i.e.,
\begin{equation}
\delta \Gamma_{WZ}(\widetilde{\phi})\ =\ -\frac{i}{4\pi}\int_\Sigma
   Tr\left(\phi^{-1}\delta\phi(\phi^{-1}d\phi)^2\right)\ =\ 0\ .
\label{variation}
\end{equation}
Thus, $\Gamma_{WZ}(\widetilde{\phi})$ will depend only on the boundary
condition (\ref{twist bc}).
We may write the Wess-Zumino term as
$\Gamma_{WZ}=\Gamma_{WZ}(U,w;\widetilde{U},\widetilde{w})$.
In the Lagrangian path integral formulation, modular invariance
is rather a trivial symmetry as long as the action (and the
measure) is well-defined on $\Sigma$.
{}For orbifold models, modular transformations can be reinterpreted
as changes of boundary conditions.
{}For example, the Wess-Zumino term should satisfy
\begin{eqnarray}
\Gamma_{WZ}(U,w;U\widetilde{U},\widetilde{w}+\widetilde{U}w) &=&
    \Gamma_{WZ}(U,w;\widetilde{U},\widetilde{w})
      \quad \mbox{mod\ $2\pi i$}\ ,
    \nonumber\\
\Gamma_{WZ}(\widetilde{U}^T,-\widetilde{U}^T\widetilde{w};U,w) &=&
    \Gamma_{WZ}(U,w;\widetilde{U},\widetilde{w})
      \quad \mbox{mod\ $2\pi i$}\ .
\label{modular tr}
\end{eqnarray}
The first (second) relation corresponds to the invariance under the
modular transformation
$T:\tau\rightarrow \tau+1 \ (S:\tau\rightarrow -1/\tau)$.
In the following, we shall explicitly express $\Gamma_{WZ}$ in terms
of $U,w,\widetilde{U}$ and $\widetilde{w}$ and verify the relations
(\ref{modular tr}).
{}Furthermore, we will see that the Wess-Zumino term can be reduced to the
$B^{IJ}$-term in eq.(\ref{action}) if both $U^{IJ}$ and
$\widetilde{U}^{IJ}$ commute with $B^{IJ}$.
To this end, we will use the Polyakov-Wiegmann formula
\cite{P-W},
\begin{equation}
\Gamma_{WZ}(\widetilde{\phi}_1 \widetilde{\phi}_2)\ =\
   \Gamma_{WZ}(\widetilde{\phi}_1) + \Gamma_{WZ}(\widetilde{\phi}_2)
   -\frac{i}{4\pi}\int_\Sigma
   Tr(\phi^{-1}_1d\phi_1 \,\phi_2d\phi^{-1}_2)\ .
\label{PWformula}
\end{equation}
In terms of the zero modes, the formula (\ref{PWformula})
may be written as
\begin{eqnarray}
\Gamma_{WZ}(U,w_1+w_2;\widetilde{U},\widetilde{w}_1+\widetilde{w}_2)
   &=& \Gamma_{WZ}(U,w_1;\widetilde{U},\widetilde{w}_1)\ +\
       \Gamma_{WZ}(U,w_2;\widetilde{U},\widetilde{w}_2) \nonumber\\
   & &\   -i\pi \bigl( w^I_1U^{IJ}\widetilde{w}^J_2
          - \widetilde{w}^I_1 \widetilde{U}^{IJ}w^J_2 \bigr)
      \quad \mbox{mod\ $2\pi i$}.
\label{PWformula2}
\end{eqnarray}
Let us write $\Gamma_{WZ}$ into the form,
\begin{eqnarray}
\Gamma_{WZ}(U,w;\widetilde{U},\widetilde{w})
   &=& i\frac{\pi}{2} w^I C^{IJ}_{\widetilde{U}}w^J
       + i\frac{\pi}{2}\widetilde{w}^I C^{IJ}_U \widetilde{w}^J
       - i\frac{\pi}{2}\widetilde{w}^I(U^TB\widetilde{U})^{IJ}w^J
       - i\frac{\pi}{2}\widetilde{w}^IB^{IJ}w^J\nonumber\\
   & &\qquad + \Delta \Gamma(U,w;\widetilde{U},\widetilde{w})\ .
\label{zeroWZW}
\end{eqnarray}
Then, it turns out that $\Delta \Gamma$ would be of the form
$\Delta \Gamma = -i2\pi(w^I\widetilde{v}^I-\widetilde{w}^Iv^I)$
modulo $2\pi i$ for some constant vectors $v^I$
and $\widetilde{v}^I$.
These constant vectors are related to shifts.
Since we are considering orbifold models without shifts, we may
have $\Delta \Gamma=0$.
The inclusion of shifts is also a topologically nontrivial
problem.
We will later discuss orbifold models associated with shifts.
It is not difficult to show that the Wess-Zumino term (\ref{zeroWZW}) with
$\Delta \Gamma=0$ satisfies the relations (\ref{modular tr}), as it should
do, and
that the one-loop vacuum amplitude (\ref{vac ampli}) exactly
agrees with the result from
the operator formalism \cite{top}.
(See eq. (6.42) in
ref.\cite{top}.)
It is now clear that the orbifold models with nontrivial twists are
quite different from those with only trivial twists in a
topological point of view because the Wess-Zumino term can be reduced to
the $B^{IJ}$-term in eq.(\ref{action}) only if both $U^{IJ}$ and
$\widetilde{U}^{IJ}$ commute with $B^{IJ}$ (then we may set
$C^{IJ}_U=C^{IJ}_{\widetilde U}=0)$.
It is worth pointing out that the antisymmetric background field
$B^{IJ}$ and even the string coordinate $X^I$ do not explicitly
appear in the Wess-Zumino term.

We finally discuss orbifold models associated with shifts.
In the construction of the orbifold models in the path
integral formalism, we might have trouble, as mentioned before.
It seems that there is no way to impose the twisted boundary
condition corresponding to the identification
$(X^I_L,X^I_R) \sim (U^{IJ}X^J_L+2\pi v^I,U^{IJ}X^J_R-2\pi v^I)$
unless
we introduce new degrees of freedom corresponding to
$\frac{1}{2}(X^I_L-X^I_R)$ besides $X^I$.
One way to introduce new degrees of freedom may be to double
the degrees of freedom and then to take the square root of the
result, as done for asymmetric orbifolds \cite{asym}.
Here we take another approach.
Our proposal is again to replace the $B^{IJ}$-term in eq.
(\ref{action}) by the Wess-Zumino term.
This time the twisted boundary condition (\ref{bc tildephi})
should be replaced by
\begin{eqnarray}
\eta(U;\alpha)\ \ &\longrightarrow&\ \
     \eta(U,v;\alpha)\ =\ \exp \left\{
       i\frac{\pi}{2}\alpha^IC^{IJ}_U\alpha^J
       -i2\pi \alpha^Iv^I\right\}
     \ ,\nonumber\\
\eta(\widetilde{U};\alpha)\ \ &\longrightarrow&\  \
     \eta(\widetilde{U},\widetilde{v};\alpha)\ =\
     \exp \left\{
       i\frac{\pi}{2}\alpha^IC^{IJ}_{\widetilde{U}}\alpha^J
       -i2\pi\alpha^I\widetilde{v}^I\right\}\ .
\label{bcshift}
\end{eqnarray}
Since the rotations are irrelevant to our present discussion,
we will set $U^{IJ}=\widetilde{U}^{IJ}=\delta^{IJ}$ for simplicity.
Then, we can show that
\begin{equation}
\Gamma_{WZ}(v,w;\widetilde{v},\widetilde{w})\ =\
    -i\pi\widetilde{w}^IB^{IJ}w^J - i2\pi(w^I\widetilde{v}^I -
     \widetilde{w}^Iv^I)\qquad \mbox{mod\ $2\pi i$}\ .
\label{shiftaction}
\end{equation}
We shall first prove eq.(\ref{shiftaction}) for
$G=SU(2)$ and then for any simply-laced Lie algebra $G$.
In the case of $SU(2)$, the string coordinate $X(\sigma^1,\sigma^2)$
propagates on a one dimensional space, i.e., the maximal torus of
the group manifold $SU(2)$.
The field $\phi$ defined in eq.(\ref{def phi}) may be given by
a $2\times 2$ matrix,
\begin{equation}
\phi(\sigma^1,\sigma^2)\ =\
\left(\matrix{
   e^{i\sqrt{2}X(\sigma^1,\sigma^2)} & 0\cr
   0 & e^{-i\sqrt{2}X(\sigma^1,\sigma^2)}\cr}
\right),
\label{phiSU(2)}
\end{equation}
in the fundamental representation of $SU(2)$.
The $X(\sigma^1,\sigma^2)$ satisfies the untwisted boundary
condition with $w,\widetilde{w} \in \sqrt{2}{\bf Z}$.
An extension from $\phi(\sigma^1,\sigma^2)$ to
$\widetilde{\phi}(t,\sigma^1,\sigma^2)$ may be given by
\begin{equation}
\widetilde{\phi}(t,\sigma^1,\sigma^2)\ =\
\left(\matrix{
    f(t)\, e^{i\sqrt{2}X(\sigma^1,\sigma^2)} &
    -\sqrt{1-f(t)^2}\, e^{-i2\sqrt{2}Y(\sigma^1,\sigma^2)}\cr
    \sqrt{1-f(t)^2}\, e^{i2\sqrt{2}Y(\sigma^1,\sigma^2)} &
    f(t)\, e^{-i\sqrt{2}X(\sigma^1,\sigma^2)}\cr}
\right),
\label{ansatz}
\end{equation}
with $f(1)=1$ and $f(0)=0$.
It turns out that $\widetilde{\phi}(t,\sigma^1,\sigma^2)$ satisfies
the desired boundary condition if the function
$Y(\sigma^1,\sigma^2)$ satisfies
\begin{eqnarray}
Y(\sigma^1+1,\sigma^2) &=&
    Y(\sigma^1,\sigma^2) + \pi v\ ,\nonumber\\
Y(\sigma^1,\sigma^2+1) &=&
    Y(\sigma^1,\sigma^2) + \pi \widetilde{v}\ .
\label{bcY}
\end{eqnarray}
We should notice that $\widetilde{\phi}$ in eq.(\ref{ansatz})
contains a new degree of freedom $Y(\sigma^1,\sigma^2)$,
which will correspond to the variable $\frac{1}{2}(X_L-X_R)$.
It is easy to see that the ansatz (\ref{ansatz}) leads to
eq.(\ref{shiftaction}).
(The antisymmetric background field vanishes in one dimension.)

In order to extend the above result to any simply-laced Lie
group $G$, let us first write $\phi(\sigma^1,\sigma^2)$ into
the form,
\begin{equation}
\phi(\sigma^1,\sigma^2)\ =\
    \phi_1(\sigma^1,\sigma^2)\,
    \phi_2(\sigma^1,\sigma^2)\cdots
    \phi_D(\sigma^1,\sigma^2)\ ,
\end{equation}
where
\begin{equation}
\phi_i(\sigma^1,\sigma^2)\ =\
    \exp \left\{ i2\mu^i\cdot X(\sigma^1,\sigma^2)\,
                 \alpha_i\cdot H \right\}\ .
\end{equation}
Here, $\mu^i$ $(i=1,...,D)$ denotes the fundamental weights
satisfying $\mu^i\cdot \alpha_j = \delta^i_{\ j}$.
An extension from $\phi(\sigma^1,\sigma^2)$ to
$\widetilde{\phi}(t,\sigma^1,\sigma^2)$ may be of the form,
\begin{equation}
\widetilde{\phi}(t,\sigma^1,\sigma^2)\ =\
    \widetilde{\phi}_1(t,\sigma^1,\sigma^2)\,
    \widetilde{\phi}_2(t,\sigma^1,\sigma^2)\cdots
    \widetilde{\phi}_D(t,\sigma^1,\sigma^2)\ .
\end{equation}
Noting that $\alpha_i\cdot H$ and $E^{\pm \alpha_i}$ form
a $SU(2)$ subgroup of $G$, we may assume that
$\widetilde{\phi}_i(t,\sigma^1,\sigma^2)$
is restricted to the subgroup $SU(2)$ whose generators consists
of $\alpha_i\cdot H$ and $E^{\pm \alpha_i}$ and
that $\widetilde{\phi}_i(t,\sigma^1,\sigma^2)$ has a similar
form to eq.(\ref{ansatz}).
Using the Polyakov-Wiegmann identity repeatedly, we find that
\begin{eqnarray}
\Gamma_{WZ}(\widetilde{\phi}) & = &
\Gamma_{WZ}(\widetilde{\phi}_1\widetilde{\phi}_2\cdots \widetilde{\phi}_D)
\nonumber\\
& = & \sum_{i=1}^D \Gamma_{WZ}(\widetilde{\phi}_i)
- \sum_{i<j} \frac{i}{4\pi}
\int_{\Sigma}Tr(\phi_i^{-1}d\phi_i\,\phi_j d\phi_j^{-1})\ .
\end{eqnarray}
The first sum is just a contribution from each of the
$SU(2)$ subgroups of $G$,
and they give rise to the term
$- i2\pi(w^I\widetilde{v}^I - \widetilde{w}^Iv^I)$ in equation
(\ref{shiftaction}), while the second term is a kind of
\lq\lq interaction term", and
can be shown to give $-i\pi\widetilde{w}^IB^{IJ}w^J$ modulo $2\pi i$.
We can then show that
the one-loop vacuum amplitude agrees with the result from the operator
formalism.

We have found that the WZW action where the field $\widetilde{\phi}$
is restricted to the Cartan subgroups of the simply-laced Lie
groups on $\Sigma$, is a natural and nontrivial extension of
the string action (\ref{action}).
We have restricted our attention to the orbifold models associated
with the simply-laced Lie algebras.
The problems addressed in this paper may not, however, be
peculiar to those models.
This suggests that we might have a generalized WZW action
which is not associated with Lie algebras.


\vspace{1cm}
\begin{center}
{\Large\bf Acknowledgements \par}
\end{center}

We should like to thank S. Fujita, T. Maskawa and J.L. Petersen
for useful discussions and H. Tsukada for drawing our attention
to ref.\cite{C-G}.
One of the authors (M.S.) would like to acknowledge the hospitality
of the Niels Bohr Institute where part of this work was done.


\end{document}